# Probing local emission properties in InGaN/GaN quantum wells by scanning tunneling luminescence microscopy

Mylène Sauty,[1] Natalia Alyabyeva,[1] Cheyenne Lynsky,[2] Yi Chao Chow,[2] Shuji Nakamura,[2] James S. Speck,[2] Yves Lassailly,[1] Alistair C. H. Rowe,[1] Claude Weisbuch,[1,2] and Jacques Peretti[1,*]

[1]Laboratoire de Physique de la Matière Condensée, Ecole polytechnique, CNRS, Institut Polytechnique de Paris, 91120 Palaiseau, France

[2]Materials Department, University of California, Santa Barbara, California 93106, USA

*e-mail: jacques.peretti@polytechnique.edu

**Abstract**

Scanning tunneling electroluminescence microscopy is performed on a 3-nm-thick InGaN/GaN quantum well with $x = 0.23$ such that the main light emission occurs in the green. The technique is used to map the local recombination properties at a scale of ~10 nm and to correlate them with the surface topography imaged by scanning tunneling microscopy. A 500 nm × 500 nm area around a 150-nm large and 2.5-nm deep hexagonal defect is probed, revealing emission at higher energies close to the defect edges, a feature which is not visible in the macro-photoluminescence spectrum of the sample. Via a fitting of the local tunneling electroluminescence spectra, quantitative information on the fluctuations of the intensity, energy, width and phonon replica intensity of the different spectral contributions are obtained, revealing information about carrier localization in the quantum well. This procedure also indicates that carrier diffusion length on the probed part of the quantum well is approximately 40 nm.

## 1. Introduction

III-Nitride heterostructures have found wide use in optoelectronic devices, in particular as energy efficient light sources.[1] However, although the nitride light emitting diodes (LEDs) made of InGaN/GaN quantum wells (QW) have outstanding performance at low injection current and low indium content, the extension of nitride light emitting devices to high current densities and long wavelengths is still challenging.[1,2] When compared to other conventional III-V semiconductors such as GaAs, the crystalline quality of nitride materials is poor, with large dislocation and defect densities. Furthermore, InGaN/GaN QWs present broad photoluminescence (PL) spectra compared to conventional semiconductors.[3] The origin of this inhomogeneous broadening is still debated and may be due to large- and small-scale composition fluctuations arising from both intrinsic alloy disorder and extrinsic sources such as clustering or large-scale variations in the indium incorporation related to defects and morphology.[4,5,6,7] For example, near field PL mapping indicates fluctuations in the emission properties linked to large scale composition fluctuations at the micron scale,[5,6,8] and underlines the link between the indium incorporation and the surface morphology in InGaN thick layers.[9] Another near-field technique based on scanning tunneling electroluminescence (STL) microscopy[10,11,12,13,14] was used to identify the origin of PL broadening in nitride materials.[15,16,17] However, experiments were performed on inappropriate heterostructures with multiple QWs or a single QW too far from the surface which led to a degraded spatial resolution, allowing only for fluctuations in the emission at large spatial scales to be observed. More recently, STL has nonetheless shown its capability to assess smaller scale sources of PL broadening.[18]

## 2. Method

In this paper, we use STL microscopy to map the optoelectronic properties of an InGaN/GaN quantum well specifically designed to permit electroluminescence imaging down to the nanometer scale. The technique relies on the use of a scanning tunneling microscope (STM) tip as a very local current source to probe the electroluminescence properties of a material at the nanometer scale and correlate light emission to the surface topography obtained simultaneously by STM.[19,20] A schematic of the experiment is shown in Figure 1(a). The STM tip is used both to image the surface topography and to locally inject electrons into the p-type QW whose details are given immediately below. The sample has a single $In_{0.23}Ga_{0.77}N$ QW located at only 30 nm from the surface, which reduces the lateral extension of the injection cone formed by the electrons that reach the QW. Electron recombination with majority holes in the p-type InGaN QW results in light emission, which is collected and analyzed by a spectrometer. Figure 1(b) shows a schematic view of the electroluminescence process in the sample bandstructure in real space.

The sample was grown by MOCVD on a sapphire substrate, followed by 2 μm of unintentionally doped (UID) GaN, 2 μm of n-doped GaN, 25 nm of UID GaN, 100 nm of p-doped GaN, 40 nm of low temperature flow-rate modulation epitaxy (FME) GaN, 250 nm of UID GaN, 3 nm of InGaN, 10 nm of UID GaN and 20 nm of p-doped GaN. The corresponding dopant concentrations are indicated in the sample structure in the inset of Figure 1(c). The active layer consists in a 3-nm-thick InGaN QW with a nominal indium concentration of 23 %, leading to green light emission. It should be noted that the STL measurement requires the presence of majority carriers, holes in this case, both to ensure sufficient conductivity for STM and to provide holes for the recombination of the electrons injected in the QW. This is achieved by a modulation doping of the QW, obtained by the growth of p-doped GaN layers on both sides of the InGaN QW, which results in an equilibrium hole concentration in the QW of about $4\times10^{12}$ cm$^{-2}$ at room temperature, without excessive nonradiative recombination that is ubiquitous in Mg-doped GaN. Very low Mg concentration in the QW is ensured by the growth, after the first p-doped GaN layer, of a sufficiently thick UID GaN layer and the 40 nm-thick FME GaN layer.[21] When excited with 3.3 eV photons, the sample PL spectrum, shown in Figure 1(c), presents a strong emission peaked at about 2.3 eV, as expected from the QW characteristics. Its full width at half maximum is about 125 meV, which is much larger than k$_B$T at 100 K (i.e. 10 meV) and is comparable to values reported in the literature.[3]

The spatial resolution of the STL measurement depends on several factors. First, the resolution of the electron injection at the surface, which can be evaluated by the quality of the topography image, is typically smaller than 1 nm. Then, the electrons form an injection cone whose angle is given by momentum conservation during the crossing of the band bending region that exists at the sample surface due to the aforementioned p-type doping. An estimation of an upper limit for this angle is π/4 radians,[10] which corresponds to a radius of injection of 30 nm in the QW. Finally, 2D lateral transport inside the QW can occur before the recombination process, with a diffusion length that depends on the properties of the QW and on the temperature. More precise information on the actual resolution of the measurement is obtained through the analysis of the experimental results as described below.

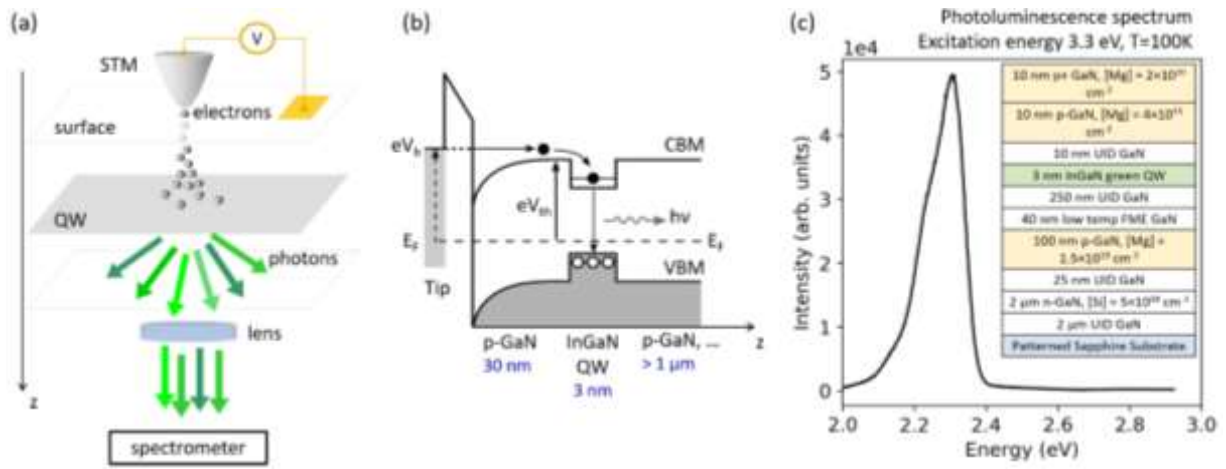

**Figure 1.** (a) Schematic of the scanning tunneling electroluminescence (STL) microscopy experiment. Electrons are locally injected from an STM tip. When reaching the InGaN QW close to the surface, they recombine with holes and emit photons which are collected in a transmission geometry and analyzed by a spectrometer. (b) Representation of the semiconductor band structure showing the conduction band minimum (CBM) and valence band maximum (VBM) in real space and of the electron injection and light emission processes during the STL experiment. The sample is p-doped, which induces a downwards band banding region close to the surface. (c) Sample photoluminescence spectrum recorded at 100 K with 3.3 eV excitation energy. The inset shows the full sample structure.

## 3. Results

The experiment was performed in an ultra-high-vacuum (UHV) STM, with chemically etched tungsten tips. A typical STM topography image taken at 100 K is shown in Figure 2(a) for a tunneling voltage of 4.4 V and a tunneling current of 0.5 nA. Note that STM on large bandgap semiconductors requires the use of severe injection conditions. A bias voltage of several volts had to be applied between the tip and the metallic contact at the sample surface, with particularly high values at low temperature due to contact and access resistance. Despite this, excellent topographic images could be obtained. The topography image in Figure 2(a) shows a structure of atomically flat terraces separated by mono- or bi-atomic steps arising from the substrate miscut. A 150 nm-wide hexagonal hole is also observed in the field of view. Two topography line profiles labelled 1 and 2 are shown below the image. Profile 1 shows the 80-nm-wide atomically flat terraces separated by atomic steps. Profile 2, which is taken through the hexagonal hole, shows that this defect is 2.5 nm deep and has a flat bottom. Note the topography profile inside the hole which indicates that the STM tip is indeed in tunneling contact at the bottom of the hole. A possible origin for this defect could be a polarity inversion region with a reversed pyramidal shape that has grown in the GaN p-type layer present below the QW, inducing different growth speeds on top of it and leading to a hexagonal flat-bottom hole in the topography.[22] Such a hole would then lead to a discontinuity of the QW position and probably to variations of the QW thickness on its edges.

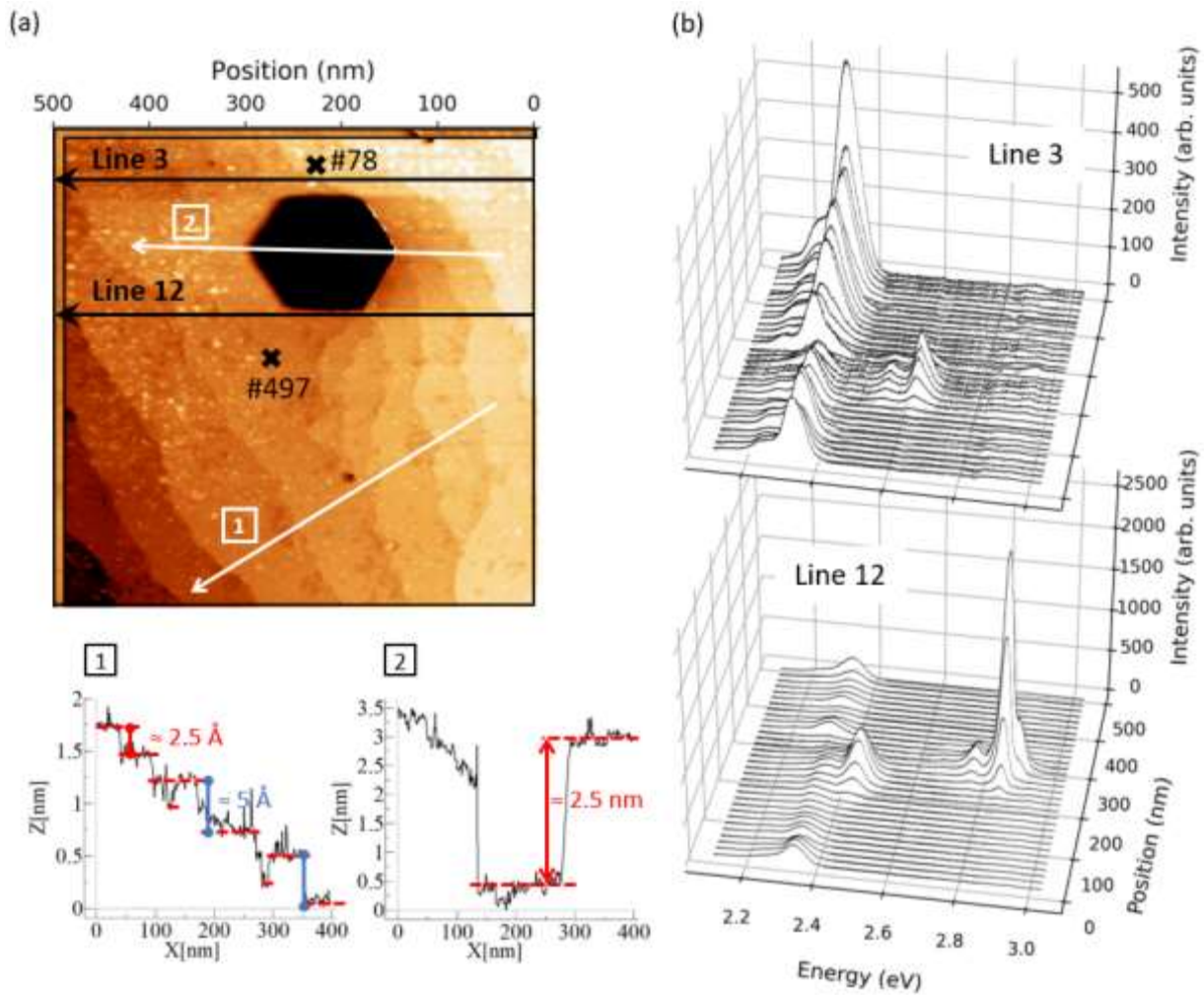

**Figure 2.** (a) Surface topography on a 500 nm × 500 nm region, that includes a large 2.5-nm-deep hexagonal flat defect, with parameters 4.4 V and 0.5 nA. Two topography profiles recorded along lines labelled 1 and 2 are shown, highlighting the monoatomic and biatomic (in blue) steps and the defect structure, respectively. (b) Sets of 32 spectra recorded with injection parameters of 4.4 V and 4 nA, along the scan lines labelled 3 and 12 indicated on the topography image. Measurements are performed at 100 K.

The electroluminescence measurement was then carried out on the same area, which was easily recognizable thanks to the wide defect. It consisted in recording the luminescence spectra on a 32 × 32 array of points regularly spaced on the 500 nm × 500 nm scanned area. The light emission was measured with a spectrometer, with a spectral resolution of 15 meV and an integration time of 10 seconds per spectrum. The tunneling current was increased to 4 nA, in order to increase the luminescence intensity. Reducing the temperature to 100 K was also essential to increase the electroluminescence signal, thanks to a strong reduction in non-radiative processes. The 32 spectra along the scan lines number 3 and 12 (indicated on the topography image) are shown in Figure 2(b). Surprisingly, in addition to the expected green emission peaked at about 2.33 eV which is observed almost over the entire sample area, several distinct, highly localized, contributions appear, including an intense blue emission peaked at about 2.85 eV.

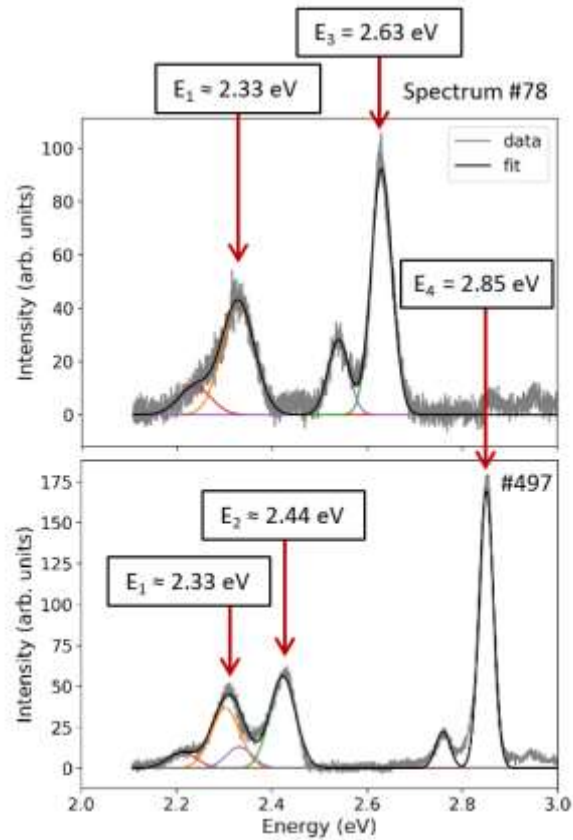

**Figure 3.** Two spectra from the dataset shown in Figure 2. Their respective positions are indicated by markers on the topography image. The different contributions and their first phonon replica are fitted by Gaussian pairs.

As an illustration, two individual spectra, numbered 78 and 497, are plotted in Figure 3. The corresponding positions where these spectra were recorded are indicated on the topography image of Figure 2(a). These selected spectra exhibit four emission peaks, respectively at 2.33 eV, 2.44 eV, 2.63 eV and 2.85 eV, which are representative of the four main emission peaks present in the scanned area. Each of them is accompanied by a phonon replica, appearing 91 meV lower in energy. To extract quantitative information on the intensity, peak energy and width of these four contributions, data fitting was performed using pairs of Gaussian functions. The two Gaussians of each pair represent a peak and its phonon replica, have identical widths and are separated by 91 meV. Respective fits for the two spectra 78 and 497 are also shown in Figure 3.

## 4. Discussion

Figure 4 shows maps of intensity, peak energy and full width at half maximum (FWHM) of the four contributions present in the scanned area, obtained by fitting all spectra as explained above. The intensity maps are plotted in logarithmic scale because of the large intensity fluctuations observed across the probed area. For comparison purposes, for each variable the scales cover identical ranges shifted when necessary. In a region where a contribution is not present, the intensity map is dark, whereas the energy and width maps appear white. The position of the hexagonal defect is indicated in all maps.

As already mentioned, the contribution at about 2.33 eV corresponds to green light emission, which is the expected emission from the 3-nm QW with 23% Indium content. It is present on the whole scanned area, and its phonon replica has a relative intensity of about 0.22, in agreement with already reported values in PL for green InGaN QW.[3] This emission corresponds to the dominant emission in the PL spectrum. However, in contrast with the PL spectrum shown in Figure 1(c), its shape is well fitted by a single Gaussian function (and its phonon replica), with a full width at half maximum of about 70 meV, nearly two times smaller than the width of the PL spectrum. Furthermore, this local emission spectrum exhibits fluctuations in intensity, energy and width both at large (100 nm) and small (a few nm) spatial scales. These fluctuations do not appear to be correlated to the surface morphology. Note that, the energy and spatial scales of the large-scale fluctuations seem compatible with measurements reported in the littérature using near-field PL mapping.[7,8]

The emission characteristics of this 2.33 eV contribution close to the hexagonal defect are also interesting. The emission energy is similar inside and outside the defect, but the spatial fluctuations of the peak energy are much smaller inside than outside the defect. Moreover, the emission energy map presents sharp edges that have the same shape as the defect and are shifted by only 20 nm in position. This indicates that there is no transport of electrons from inside to outside the defect and that the defect exists nearly exactly at the same spatial position in the QW as at the sample surface. This is compatible with the aforementioned possible origin of this hexagonal defect as being related to a polarity inversion region which would isolate the QW in the defect area from the surrounding QW area.

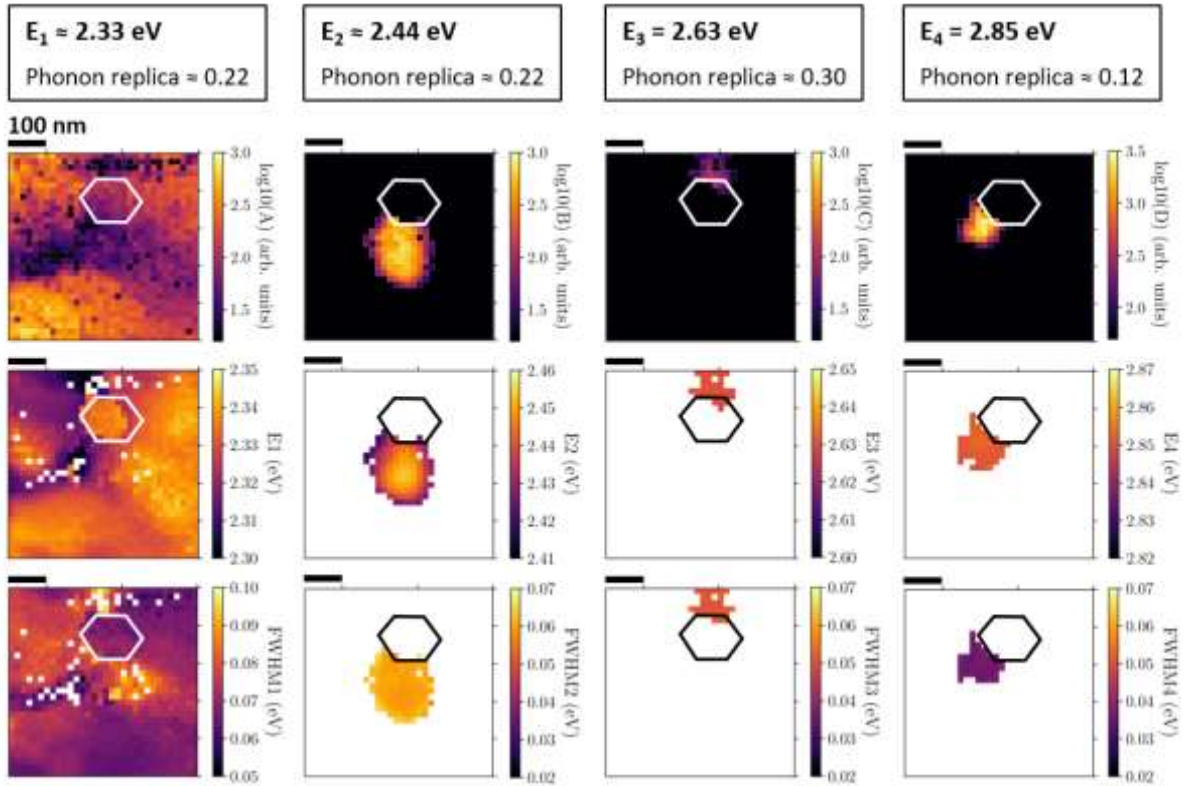

**Figure 4.** Maps of intensity, energy and full width at half maximum (FWHM) for the four contributions appearing in the dataset of Figure 2(b). The average value of the phonon replica relative intensity is given for each contribution. The position of the hexagonal defect seen in the topography images is indicated in all maps. The 3 high energy contributions appear only on small regions located close to edges of the hexagonal defect. In the intensity maps, black areas correspond to regions where the corresponding contribution is not observed, whereas in the energy and width maps these regions appear in white.

The three other contributions are very surprising since they correspond to emission at much higher energies than the green emission. Moreover, they can be very intense and spectrally narrow. Finally, as shown in their intensity, peak energy and FWHM, they are observed in regions of small spatial extension (about 100 nm) near the hexagonal defect edges. However, although these contributions can be very intense in the local electroluminescence spectra, they are not visible in the macroscopic PL measurement presented in Figure 1, probably because they represent only a small fraction of the total emitted light, arising from a small fraction of emitting area.

We can distinguish two different behaviors. Contributions at 2.63 and 2.85 eV both appear at a fixed energy and width. They most probably arise from a modification of the QW on the edges of the defect, due either to a thinning of the QW (2.63 and 2.85 eV would correspond to QW thicknesses 2nm and 1.25nm, respectively) or to a lower indium incorporation. When injection occurs outside the defect, the electron states leading to this high energy emission are accessible from a distance of about 80 nm. This gives an estimation of the electron diffusion length. This value is in agreement with reported data in the literature.[7] However, when injection occurs inside the defect, there is no diffusion to these states, which might be because the QW in the defect area is isolated from the surroundings. A surprising feature is that both peaks have very different phonon replica intensity, which suggests that they correspond to electronic states with very different localization lengths.[3]

In contrast to this, the 2.44 eV contribution which appears along the bottom edge of the defect shows a significant gradient in emission energy over a few tens of meV. Furthermore, the central point of this gradient is positioned at about 100 nm from the defect edge. This additional contribution most probably does not originate from a fluctuation in QW width at the defect edge. Rather, it might be due to a gallium-rich region, whose formation could have been induced by the presence of the defect. Note that in all the regions where high energy emission occurs, the main green emission is most often still present but less intense (Figure 4, left), which means that whatever the injection position, some of the injected carrier diffuse to green regions.

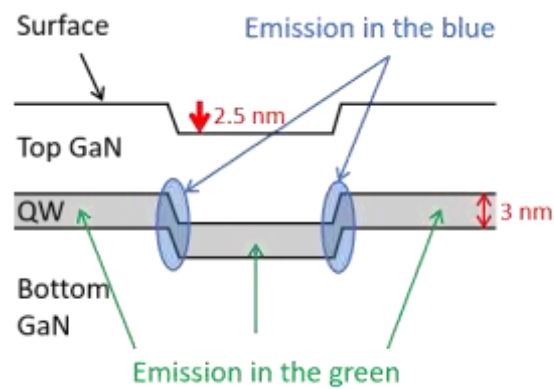

**Figure 5.** Schematic of the QW structure close to the hexagonal defect that can be deduced from the combination of the STM topography measurement and the light emission properties measured by STL microscopy.

Finally, the electroluminescence map can be used to extract a lot of information about the QW structure close to the defect. Indeed, it shows that the main light emission inside and outide the defect is in the green, and corresponds to the nominal QW thickness of 3 nm and indium concentration of 23 %. On the edges of the defect, states higher in energy exist, which lead to an emission in the blue. Injected electrons can diffuse to these states only when they are injected outside the defect. Figure 5 shows a schematic of the QW structure close to the defect that can be deduced from the combination of the topography measurement and the light emission properties.

## 5. Conclusion

In this work, we show that STL microscopy can be used to image and quantitatively analyze the local recombination properties of an InGaN QW down to the nanometer scale. We can distinguish large- and small-scale fluctuations in emission, and correlate them with the surface topography. The recombination properties around defects can be investigated in detail, as is done close to a hexagonal defect with a flat bottom visible in the surface topography. The resolution of the measurement is limited by the transport of carriers inside the material, which can be strongly position dependent, as is shown with the injection of carriers inside or outside the defect. The imaging of spatially localized states with specific emission properties allows to estimate a local carrier diffusion length of around 40 nm in the QW. Finally, the fit of the local emission spectra is used to precisely extract the emission properties of each contribution such as their intensity, peak energy, width and the relative intensity of the phonon replica. The spatial variations of these emission properties can give insights into the nature of the local electronic states. In the data presented here, it is possible to distinguish two different sources for the high energy contributions, depending whether they showed a gradient in the emission energy or not. Finally, such measurement and data treatment allows to get insight on the nature of the light emitting region, and its structure close to the imaged defect.

The spatially controlled carrier injection permitted by the STL technique opens the way to unique possibilities for the observation of the recombination properties in semiconductors at the local scale, under various configurations and samples structures. In is particularly relevant for the identification of defects and for the study of their optoelectronical properties, which is of much interest in the growth point of view. Finally, the possibility to image the surface topography at the same time as the nanoscale electroluminescence can be used to unambiguously distinguish fluctuations in emission arising from extrinsic sources such as defects, step edges, or large-scale variations in the indium incorporation, or from intrinsic sources such as alloy disorder.

**Acknowledgements**

We thank Jean-Philippe Banon for fruitful discussion and Erwan Allys for precious help with data processing. This work was supported by the French National Research Agency (ANR, grants ELENID No. ANR-17-CE24-0040-01 and TECCLON No. ANR-20-CE05-0037-01). JSS and CW are supported by the Simons Foundation (Grants No 601952 and 601954, respectively) and by the National Science Foundation (NSF) RAISE program (Grant No. DMS-1839077).